\documentclass[dvipdfm]{sig-alternate}

\usepackage{graphicx}
\usepackage{url}
\usepackage{xspace}
\usepackage{amsfonts}

\usepackage{color}
\definecolor{mycolor}{rgb}{1,0,0}

\newcommand{\comment}[1]{}

\newcommand{\tw}{trustworthiness-weighted\xspace}

\def\thepage{\arabic{page}}
\begin{document}
%
\CopyrightYear{2012} 

\title{A Framework of Algorithms: Computing the Bias and Prestige of Nodes in Trust
Networks}



%
%
%
%
%

%
\author{
%
%
 Rong-Hua Li \quad
%
Jeffrey Xu Yu \quad
%
Xin Huang \quad
Hong Cheng \\
\affaddr{The Chinese University of Hong Kong} \\
\email{\{rhli,yu,xhuang,hcheng\}@se.cuhk.edu.hk }}

\newcounter{example}[section]
\renewcommand{\theexample}{\nthesection.\arabic{example}}
\newenvironment{example}{
     \refstepcounter{example}
     {\vspace{1ex} \noindent\bf  Example  \theexample:}}{
     \eop\vspace{1ex}} 

\newcounter{definition}[section]
\renewcommand{\thedefinition}{\nthesection.\arabic{definition}}
\newenvironment{definition}{
     \refstepcounter{definition}
     {\vspace{1ex} \noindent\bf  Definition  \thedefinition:}}

\newcounter{theorem}[section]
\renewcommand{\thetheorem}{\nthesection.\arabic{theorem}}
\newenvironment{theorem}{\begin{em}
        \refstepcounter{theorem}
        {\vspace{1ex} \noindent\bf  Theorem  \thetheorem:}}{
        \end{em}}

\newcounter{lemma}[section]
\renewcommand{\thelemma}{\nthesection.\arabic{lemma}}
\newenvironment{lemma}{\begin{em}
        \refstepcounter{lemma}
        {\vspace{1ex}\noindent\bf Lemma \thelemma:}}{
        \end{em}}

\newcounter{corollary}[section]
\renewcommand{\thecorollary}{\nthesection.\arabic{lemma}}
\newenvironment{corollary}{\begin{em}
        \refstepcounter{corollary}
        {\vspace{1ex}\noindent\bf Corollary \thecorollary:}}{
        \end{em}}

\newcounter{remark}[section]
\renewcommand{\theremark}{\thesection.\arabic{remark}}
\newenvironment{remark}{\begin{em}
        \refstepcounter{remark}
        {\vspace{1ex}\noindent\bf Remark \theremark:}}{
        \end{em}\eop\vspace{1ex}} 

\newcommand{\proofsketch}{\noindent{\bf Proof Sketch: }}
\newcommand{\myproof}{\noindent{\bf Proof: }}

\newcommand{\nthesection}{\arabic{section}}

\newcommand{\eop}{\hspace*{\fill}\mbox{$\Box$}}

\newcommand{\stitle}[1]{\vspace{1ex} \noindent{\bf #1}}

\newcommand{\kw}[1]{{\ensuremath {\mathsf{#1}}}\xspace}



\maketitle

\begin{abstract}
A trust network is a social network in which edges represent the
trust relationship between two nodes in the network. In a trust
network, a fundamental question is how to assess and compute the
bias and prestige of the nodes, where the bias of a node measures
the trustworthiness of a node and the prestige of a node measures
the importance of the node. The larger bias of a node implies the
lower trustworthiness of the node, and the larger prestige of a node
implies the higher importance of the node. In this paper, we define
a vector-valued contractive function to characterize the bias vector
which results in a rich family of bias measurements, and we propose
a framework of algorithms for computing the bias and prestige of
nodes in trust networks. Based on our framework, we develop four
algorithms that can calculate the bias and prestige of nodes
effectively and robustly.  The time and space complexities of all
our algorithms are linear w.r.t.\ the size of the graph, thus our
algorithms are scalable to handle large datasets. We evaluate our
algorithms using five real datasets. The experimental results
demonstrate the effectiveness, robustness, and scalability of our
algorithms.
\end{abstract}





\vspace*{-0.2cm}
\section{Introduction}
\label{sec:intro} 

Online social networks (OSNs) such as Facebook, Twitter, and MySpace
have become increasingly popular in recent years. These OSNs provide
users with various utilities to share all sorts of information with
friends, such as thoughts, activities, photos, etc.  As a new way to
express information, trust networks such as Advogato
(\url{www.advogato.org}), Kaitiaki (\url{www.kaitiaki.org.nz}),
Epinions (\url{www.epinions.com}), and Slashdot
(\url{www.slashdot.org}) rapidly attract more and more attention.
Unlike the traditional OSNs where edges represent the friendship
between users, in the trust networks, edges express the trust
relationship between two users. In other words, users express their
trust to other users by giving a trust score to another, and users are
evaluated by others based on their trust scores.
There exist two types of trust networks, namely, unsigned and
signed. In the unsigned trust networks, such as Advogato and Kaitiaki,
users can only express their trust to other users by giving a
non-negative trust score to others. In the signed trust networks, such
as Epinions and Slashdot, users can express their trust or distrust to
others by giving a positive or negative trust score to others.  There are many applications in the trust networks, such as finding
the trusted nodes in a network \cite{11wwwbiasrank}, predicting the
trust score of the nodes, and the trust based recommendation systems
\cite{07trustrs}.

In a signed/unsigned trust network, the final trustworthiness of a
user is determined by how users trust each other in a global
context, and is measured by \emph{bias}.  The bias of a user
reflects the extent up to which his/her opinions differ from others.
If a user has a zero bias, then his/her opinions are 100\% unbiased
and can be 100\% taken. Consequently, the user has high
trustworthiness. On the other hand, if a user has a large bias, then
his/her opinions cannot be 100\% taken because his/her opinions are
often different from others. Therefore, the user has low
trustworthiness. Another important measure, the \emph{prestige} of a
user, reflects how he/she is trusted by others (the importance).  In
this work, we study how to assess and compute the bias and prestige
of the users. The challenges are: (1) how to define a reasonable
bias measurement that can capture the bias of the users' opinions,
(2) how to handle the negative trust scores in signed trust
networks, and (3) how to design a robust algorithm that can prevent
attack from some adversarial users.

As pointed out in \cite{11wwwbiasrank}, the traditional eigenvector
based methods, such as eigenvector centrality
\cite{72eigenvectorcentrality}, HITS \cite{pagerank}, and PageRank
\cite{99jacmhits}, cannot be directly used to solve this problem. The
reason is of twofold. First, the eigenvector based methods typically
cannot handle the negative edges. Moreover, they cannot distinguish
the two cases, namely non-connection and zero trust score, where a
zero trust score implies that there is an edge from a user to another
user with the zero trust score (e.g., $W_{32}$ in
Fig.~\ref{fig:example}). Second, they ignore the bias of the nodes.
To the best of our knowledge, the algorithm proposed by Mishra and
Bhattacharya \cite{11wwwbiasrank} is the only algorithm that addresses
this problem. We refer to this algorithm as the MB algorithm (or
simply MB).  MB is tailored for the signed trust networks, and can
also be used for the unsigned trust networks.  However, MB has major
drawbacks. The trustworthiness of a user cannot be trusted due to the
fact that MB treats bias of a user by relative differences between
itself and others. For instance, if a user gives all his/her friends a
much higher trust score than the average of others, and gives all
his/her foes a much lower trust score than the average of others, such
differences cancel out, which leads to a zero bias for the user. This
cancellation happens in either a signed or a unsigned trust
network. Therefore, MB can be attacked by the adversarial users.  We
will analyze it in Section \ref{sec:algorithms} in detail.

In this paper, we propose new bias measurements to capture the bias
of the users' opinions. First, we define a vector-valued contractive
function as a framework to represent the bias vector, which implies
a rich family of bias measurements and thereby results in a rich
family of algorithms. On the basis of our framework, we develop four
new bias measurements using absoluate differences instead of
relative differences to deal with bias, in order to avoid such a
cancellation problem in MB. Based on the bias of the nodes, the
trustworthiness score of a node is inversely proportional to the
bias score of the node, and the prestige of a node is the average
\tw trust scores. In other words, if a node is with a large bias
score, then the trust scores given by this node will be assigned to
small weights. Our algorithms iteratively refine the bias and
prestige scores of the nodes. The final bias and prestige vector is
obtained when the algorithm converges. The main contributions of
this paper are summarized as follows.

 \begin{itemize}
 \item
First, we propose a framework of algorithms for computing the bias
and prestige of the nodes in either unsigned or signed trust
networks. We rigorously prove the convergence properties of our
framework.
 \item
Second, based on the proposed framework, we show that MB
\cite{11wwwbiasrank} is a special case of our framework for unsigned
trust networks. We also develop four new algorithms that can
overcome the cancellation problem in MB. The bias measurements of
our new algorithms are more reasonable, and our algorithms are more
effective and more robust than MB.
\item
Third, we conducted extensive experimental studies to confirm the
effectiveness, robustness, and scalability of the proposed
algorithms. We compare our algorithms with the state-of-the-art
algorithm (MB) over five real datasets. The results indicate that
our newly proposed algorithms outperform MB in terms of both
effectiveness and robustness. We also evaluate the scalability of
the proposed algorithms, and the results demonstrate the scalability
of our algorithms is linear w.r.t.\ the size of the graphs.
 \end{itemize}

The rest of this paper is organized as follows. We introduce the
related work in Section \ref{sec:rlwork}. We give the preliminaries
and problem statement, and discuss the drawbacks of the existing
algorithms in Section \ref{sec:algorithms}. We propose and analyze the algorithm
framework as well as four novel specific algorithms in
Section \ref{sec:propmethod}. Extensive experimental
results are presented in Section \ref{sec:experiments}. Finally, we conclude this
work in Section \ref{sec:concl}.

\vspace*{-0.2cm}
\section{Related work}
\label{sec:rlwork} 

Our work is closely related to graph-based ranking algorithms. In
the last decades, ranking nodes in networks has attracted much
attention in both research and industry communities. There are a
large number of algorithms, such as HITS \cite{99jacmhits}, PageRank
\cite{pagerank} and its variants \cite{02wwwtopicpagerank,
03wwwppv}. Most of these ranking algorithms are based on the
spectral of the adjacency matrix of the network. A survey on using
spectral techniques for ranking can be found in
\cite{11spectralrank}. However, all of these algorithms only focus
on finding the prestige of the nodes in the networks, and do not
consider the bias of the nodes, thus cannot be directly used in our
problem. In addition, the convergence of these algorithms is
guaranteed by the non-negative matrix theory \cite{06bookpagerank}.
Therefore, they cannot be directly generalized to the signed trust
networks \cite{10chisignedgraph}, where the edges are possibly
associated with a negative weight. A straightforward method is to
remove the negative edges, and then perform the ranking algorithms
on the non-negative graph. Obviously, this method ignores the
negative relationship between the nodes of the graph, and may result
in poor ranking performance \cite{04wwwtrustpropagation}. To address
this issue, de Kerchove et al.\ \cite{08sdmpagetrust} propose a
Pagetrust algorithm, which can handle the negative edges, but the
convergence of their algorithm cannot be guaranteed.

Our work is also related to the trust management \cite{07trustsurvey,
  07powertrust, 09reputationchallange, 09gossiptrust}. Richardson et al.\ in
\cite{03trustmanagement} propose an eigenvector based algorithm for
the trust management in semantic web. Independent to Richardson's
work, Kamvar et al.\ in \cite{03wwwp2ptrust} present a similar
eigenvector based algorithm, namely Eigentrust, for the trust
management in P2P networks. Guha et al.\ in
\cite{04wwwtrustpropagation} study the problem of propagation of
trust and distrust in the networks.  Subsequently, Theodorakopoulos
et al.\ in \cite{06jsactrustalgebra} study the trust model and trust
evaluation metrics from an algebra viewpoint. They use semiring to
express a trust model and then model the trust evaluation problem as
a path problem on a directed graph.  Recently, Andersen et al.\ in
\cite{08wwwtrustrankaxiom} propose an axiomatic approach for the
trust based recommendation systems \cite{07trustrs}. All of these
algorithms are based on the spectral techniques, and cannot be
easily generalized to the signed networks. There also exist trust
based ranking algorithms \cite{10eplrankcompare}. These algorithms
are designed based on the same idea ``the users whose opinions often
differ from the others' opinions will be assigned to less trust
scores''. In the literature, such algorithms include
\cite{03mizzrank, 06eplir,
  06phyair, 10siammaareputation, 11eplrobusttrustrank}. All of these
algorithms are tailored for the bipartite rating networks, and
generalization to the general networks would be meaningless. Moreover,
all of them except \cite{10siammaareputation} cannot guarantee the
convergence. As an exception, in \cite{10siammaareputation}, de
Kerchove and Dooren prove the convergence of their algorithm, but the
rate of the convergence is q-linear. In addition, the convergence
property of their algorithm \cite{10siammaareputation} is dependent on
the decay constant, which is very hard to be determined in practice.

\vspace*{-0.2cm}
\section{Preliminaries}
\label{sec:algorithms}

We model a trust network as a directed weighted graph $G=(V, E, W)$
with $n$ nodes and $m$ edges, where $V$ represents the node set, $E$
denotes the edge set, and $W$ denotes the weights. In graph $G$, a
weight $W_{ij}$ signifies a trust score from node $i$ to node $j$. All
trust scores are normalized in the range of $[0,1]$. For simplicity,
in the following discussions, we focus on an unsigned trust network
assuming that all edge-weights are non-negative. Our approaches can be
readily generalized to signed trust networks, and we will discuss it
in Section \ref{subsec:genesignednet}.

\begin{figure}[t]
\begin{center}
\includegraphics[scale=0.5]{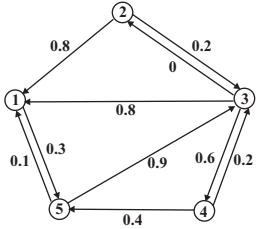}
\vspace*{-1em} \caption{A trust network.} \label{fig:example}
\end{center}
\vspace*{-0.6cm}
\end{figure}

An example is shown in Fig.~\ref{fig:example}. In
Fig.~\ref{fig:example}, node 5 gives a trust score 0.1 to node 1
($W_{51}=0.1$), whereas two nodes, 2 and 3, give a high trust score
0.8 to node 1 ($W_{21}=W_{31}=0.8$). And node 5 gives a trust score
0.9 to node 3 ($W_{53} = 0.9$), while two nodes, 2 and 4, give a
low trust score to node 3 instead ($W_{23} = W_{43} = 0.2$). This
observation shows that node 5's opinions often differ from those of
others, thus indicates that node 5 is a biased node. On the other
hand, there are two nodes (2 and 3) giving a high trust score 0.8 to
node 1 ($W_{21}=W_{31}=0.8$), which suggests that node 1 would be a
prestigious node. Additionally, in this example, node 3 gives 0 to
node 2 ($W_{32} = 0$), which implies that node 3 does not trust node
2 at all.

Given a trust network $G$, the problem we study is how to compute the
bias and prestige of the nodes based on the trust scores. As
discussed, the eigenvector based methods are not applicable, and the
only existing solution is MB \cite{11wwwbiasrank}. In the following,
we briefly review MB and discuss its major drawbacks.

In MB, each node has two scores: the bias and prestige score. The bias
and prestige scores of node $i$ are denoted by $b_i$ and $r_i$,
respectively. Formally, the bias of node $i$ is defined by
\begin{equation}
  \label{def:mbbias}
  b_i  = \frac{1}{{2|O_i |}}\sum\limits_{j \in O_i } {(W_{ij}  - r_j
  )},
\end{equation}
where $O_i$ denotes the set of all outgoing neighbors of node $i$.
The idea behind is that a node will be assigned to a high bias score if
it often behaves differently from others. The prestige score of node
$i$ ($r_i$) is given by
\begin{equation}
  \label{def:mbprestige}
  r_i  = \frac{1}{{|I_i |}}\sum\limits_{j \in I_i } {(W_{ji} (1 - \max
    \{ 0,b_j  \times sign(W_{ji} )\}
  ))},
\end{equation}
where $I_i$ denotes the set of all incoming neighbors of node $i$,
and $sign(W_{ji})$ denotes the sign of an edge from node $j$ to node
$i$, which can be positive (trust) or negative (distrust).

The MB algorithm works in an iterative fashion, and the
corresponding iterative system is

{\small
\begin{equation}   \label{eq:mbitersys}
\left\{ \begin{array}{l}
 r_i^{k+1}  = \frac{1}{{|I_i |}}\sum\limits_{j \in I_i } {(W_{ji} (1 -
   \max \{ 0,b_j^k  \times sign(W_{ji} )\} ))}  \\
 b_j^{k + 1}  = \frac{1}{{2|O_j |}}\sum\limits_{i \in O_j } {(W_{ji}
   - r_i^{k + 1} )}  \\
 \end{array} \right.
\end{equation}
}

There are two major drawbacks in MB. First, in Eq.~(\ref{def:mbbias}),
the differences ($W_{ij}-r_j$) for different outgoing neighbors $j\in
O_i$ can be canceled out, thus will result in unreasonable bias
measures. Reconsider the example (Fig.~\ref{fig:example}), node 5
gives 0.1 to node 1, while both node 2 and node 3 give 0.8 to node
1. With these three edges ($5 \rightarrow 1$, $2 \rightarrow 1$, and
$3 \rightarrow 1$), the trust score given by node 5 is significantly
lower than those of others with a difference $0.1-0.8 =
-0.7$. However, consider the other three edges $2 \rightarrow 3$, $4
\rightarrow 3$, and $5 \rightarrow 3$, we can find that the trust
score given by node 5 is significantly larger than those of the other
two nodes (nodes 2 and 4) with a difference 0.7. The positive and
negative differences can be canceled out by Eq.~(\ref{def:mbbias}),
and this will cause node 5 to be trusted with a lower bias
score. However, intuitively, node 5's opinions often differ from those
of others, thereby it should be assigned to a large bias
score. Table~\ref{tbl:mbbiasexample} shows the bias scores by MB after
each iteration. We can clearly see that node 5 gets the minimal bias
scores (0.014), which contradicts to the intuition.

\begin{table}[t]\vspace*{-0.2cm}
\begin{center}
\caption{Bias scores by the MB algorithm.} {\small
\begin{tabular}{|l|c|c|c|c|c|}
\hline
Iteration & node 1 & node 2 & node 3 & node 4 & node 5 \\
\hline
1 & 0.350 & 0.042 & 0.121 & 0.250 & 0.042 \\
2 & 0.350 & 0.015 & 0.129 & 0.232 & 0.015 \\
3 & 0.350 & 0.014 & 0.129 & 0.231 & 0.014 \\
4 & 0.350 & 0.014 & 0.129 & 0.231 & 0.014 \\
\hline
\end{tabular}
} \label{tbl:mbbiasexample}
\end{center}\vspace*{-0.8cm}
\end{table}

Second, as also pointed in \cite{11wwwbiasrank}, MB is easy to be
attacked by the adversarial nodes. For example, some nodes can
maintain their bias scores closely to 0 by giving high trust scores
to the nodes with low prestige scores and giving the low trust
scores to the nodes with high prestige scores (as node 5 in
Fig.~\ref{fig:example}). In \cite{11wwwbiasrank}, Mishra and
Bhattacharya present a statistical method for detecting such
adversarial nodes. But the statistical method is independent to MB,
thus it cannot reduce the influence of the adversarial nodes in MB.
In addition, the proof for the convergence of the MB presented in
\cite{11wwwbiasrank} is not rigorous. In the present paper, we
rigorously prove the convergence of our framework using the Cauthy
convergence theorem \cite{74analysis}.

\vspace*{-0.2cm}
\section{Our New Approach}
\label{sec:propmethod}

In this section, we propose a framework of algorithms for computing
the bias and prestige of the nodes in trust networks. In our
framework, every node $i$ has two scores: the bias score ($b_i$) and
the prestige score ($r_i$). We use two vectors $b$ and $r$ to denote
the bias vector and prestige vector, respectively.  Specifically, we
define the bias of node $j$ by
\begin{equation}
  \label{eq:trustdef}
  b_j  = (f(r))_j,
\end{equation}
where $r$ is the prestige vector of the nodes, $f(r):\mathbb{R}^n \to
\mathbb{R}^n $ is a \emph{vector-valued contractive function}, which
is defined in Definition \ref{def:vvcdef}, and $(f(r))_j$ denotes the
$j$-th element of vector $f(r)$. We restrict $0 \le f(r) \le e$, where
$e \in \mathbb{R}^n$ and $e = [1, 1, \cdots, 1]^T$.

\begin{definition}
\label{def:vvcdef} For any $x, y \in \mathbb{R}^n$, the function
$f:\mathbb{R}^n  \to \mathbb{R}^n$ is a vector-valued contractive
function if the following condition holds.
\begin{equation} \label{eq:vvccond}
|f(x) - f(y)| \le \lambda ||x - y||_\infty  e
\end{equation}
where $\lambda \in [0,1)$, $|| \bullet ||_\infty  $ denotes the
infinity norm.
\end{definition}

Since $\lambda \in [0,1)$, the vector-valued function $f$ exhibits
contractive property w.r.t.\ the infinity norm of the vector, we refer
to it as the vector-valued contractive function. It is worth noting
that the vector-valued contractive function we define is a
generalization of the contraction mapping in the fixed point theory
\cite{03fixedpoint}. In \cite{03fixedpoint}, the contraction mapping
is defined on a 1-dimensional variable and the domain of the
contraction mapping is also a 1-dimensional value.  Our vector-valued
contractive function is defined on an $n$-dimensional vector and its
domain is also an $n$-dimensional vector. The contraction mapping is
very useful for iterative function systems \cite{03fixedpoint}.
Our vector-valued contractive function sheds light on studying
the iterative vector-valued function systems in trust networks.
%
%

As can be seen in Eq.~(\ref{eq:trustdef}), the bias vector $b$ is
obtained by a vector-valued contractive function defined on the
prestige vector $r$. The advantage of the definition of bias is that
it makes our framework general, which will result in a rich family of
bias measurements. In Section \ref{subsec:specialinstance}, we will
give four different bias measurements and each of these measurements is
shown to be a vector-valued contractive function.

With the bias of the nodes, the trustworthiness of node $j$ is given
by $1-b_j$, which is inversely proportional to the bias score of
node $j$. We compute the prestige score of node $i$ by averaging the
\tw trust scores given by the incoming neighbors of node $i$. In
particular, the prestige score $r_i$ for a node $i$ is given by

\begin{equation}
\label{eq:prestigedef} r_i  = \frac{1}{{|I_i| }}\sum\limits_{j \in
I_i } {W_{ji} (1 - (f(r))_j)},
\end{equation}

\noindent
where $I_i$ is the set of all incoming neighbors of node $i$. Our
algorithm iteratively refines the prestige vector and the bias
vector using the following iterative system:

\begin{small}
\begin{equation}\label{eq:rankiter}
  \left\{ \begin{array}{l}
 r_i^{k + 1}  = \frac{1}{{|I_i| }}\sum\limits_{j \in I_i } {W_{ji} (1
   - b_j^k )}  \\
 b_j^{k + 1}  = (f(r^{k + 1} ))_j  \\
 \end{array} \right.
\end{equation}
\end{small}
\vspace*{-0.2cm}

\noindent
where $r_i^{k+1}$ denotes the prestige of node $i$ in the $(k$+$1)$-th
iteration and $b_j^{k+1}$ denotes the bias of node $j$ in the
$(k$+$1)$-th iteration. Initially, we set $f(r ^0) = 0$, which implies
$0 \le r^k \le 1$. The iterative system (Eq.~(\ref{eq:rankiter})) will
converge into a unique fixed prestige and bias vector in an
exponential rate of convergence, as we will show in Section
\ref{subsec:analysis}.

\subsection{Analysis of the proposed algorithm}
\label{subsec:analysis}

\stitle{Convergence of the proposed algorithm:} We analyze the
convergence properties of the iterative system described in
Eq.~(\ref{eq:rankiter}). Specifically, we show the prestige vector
will converge into a unique fixed point as stated in Theorem
\ref{thm:converge}. Similar arguments can be used to prove the bias
vector also converges into a unique fixed point. First, we prove the
following lemma.

\begin{lemma}
\label{lem:converge} For any node $i$, $|r_i^{k + 1}  -
r_i^k | \le \lambda ^k ||r^1  - r^0 ||_\infty$.
\end{lemma}

\begin{myproof}
We prove Lemma~\ref{lem:converge} by induction. Let $k=1$, we have

 \[
  \begin{array}{l}
 |r_i^2  - r_i^1 |= |\frac{1}{{|I_i| }}\sum\limits_{j \in I_i} {W_{ji} ((f(r^0 ))_j - (f(r^1 ))_j)} | \\
  \quad \quad \quad \; \; \; \le \frac{1}{{|I_i| }}\sum\limits_{j \in I_i} {W_{ji} |(f(r^0 ))_j - (f(r^1 ))_j|}  \\
  \quad \quad \quad \; \; \; \le \frac{\lambda }{{|I_i|}}\sum\limits_{j \in I_i} {W_{ji} ||r^1  - r^0 ||_\infty  }  \\
  \quad \quad \quad \; \; \; \le \lambda ||r^1  - r^0 ||_\infty,   \\
 \end{array}
 \]

\noindent where the second inequality is due to the definition of
vector-valued contractive function, and the last inequality is by
$|W_{ij}| \in [0,1]$. Assume the lemma holds when $k=t$. We show that
the lemma still holds when $k=t+1$.
%
\[
 \begin{array}{l}
 |r_i^{t + 2}  - r_i^{t + 1} | = |\frac{1}{{|I_i| }}\sum\limits_{j \in I_i} {W_{ji} ((f(r^t ))_j - (f(r^{t + 1}))_j)} | \\
  \quad \quad \quad \quad \quad \;\;\; \le \frac{1}{{|I_i| }}\sum\limits_{j \in I_i} {W_{ji} |(f(r^t))_j - (f(r^{t + 1}))_j|}  \\
  \quad \quad \quad \quad \quad \;\;\; \le \frac{\lambda }{{|I_i| }}\sum\limits_{j \in I_i} {W_{ji} ||r^{t + 1}  - r^t ||_\infty  }  \\
  \quad \quad \quad \quad \quad \;\;\; \le \lambda ||r^{t + 1}  - r^t ||_\infty   \\
  \quad \quad \quad \quad \quad \;\;\; \le \lambda ^{t + 1} ||r^1  - r^0 ||_\infty,   \\
 \end{array}
 \]

 \noindent where the second inequality is due to the definition of vector-valued contractive function and the last
 inequality holds by the induction assumption. This completes the proof. \eop
\end{myproof}

With Lemma \ref{lem:converge}, we prove the convergence property.

\begin{theorem}
\label{thm:converge} The iterative system defined in
Eq.~(\ref{eq:rankiter}) converges into a unique fixed point.
\end{theorem}

\begin{myproof}
We first prove the convergence of the iterative system (Eq.~(\ref{eq:rankiter})), and then prove the uniqueness.
Specifically, for $\varepsilon > 0$, there exists $N$ such that

\[
\lambda ^N  < \frac{{(1 - \lambda )\varepsilon }}{{||r^1  - r^0
||_\infty  }}.
\]

\noindent
Then, for any $s>t \ge N$, we have

\comment{
\vspace*{-0.2cm}
\begin{equation}
\[
\begin{array}{l}
 |r_i ^s  - r_i ^t | \le |r_i ^s  - r_i ^{s - 1} | + |r_i ^{s - 1}  - r_i ^{s - 2} | +  \cdots  + |r_i ^{t + 1}  - r_i ^t | \\
  \quad \quad \quad \; \; \; \le \lambda ^{s - 1} ||r^1  - r^0 ||_\infty   + \lambda ^{s - 2} ||r^1  - r^0 ||_\infty   +  \cdots
  +\\
  \quad \quad \quad \quad \; \; \; \lambda ^t ||r^1  - r^0 ||_\infty   \\
  \quad \quad \quad \; \; \;\le ||r^1  - r^0 ||_\infty  \lambda ^t \sum\limits_{k = 0}^{s - t - 1} {\lambda ^k }  \\
  \quad \quad \quad \; \; \; < ||r^1  - r^0 ||_\infty  \lambda ^t \sum\limits_{k = 0}^\infty  {\lambda ^k }  \\
  \quad \quad \quad \; \; \; = ||r^1  - r^0 ||_\infty  \lambda ^t \frac{1}{{1 - \lambda }} \\
  \quad \quad \quad \; \; \; \le ||r^1  - r^0 ||_\infty  \lambda ^N \frac{1}{{1 - \lambda }} \le \varepsilon,  \\
 \end{array}
 \]
\end{equation}
\vspace*{-0.2cm}
}

\[
\begin{array}{l}
 |r_i ^s  - r_i ^t | \le |r_i ^s  - r_i ^{s - 1} | + |r_i ^{s - 1}  - r_i ^{s - 2} | +  \cdots  + |r_i ^{t + 1}  - r_i ^t | \\
  \quad \quad \quad \; \; \; \le \lambda ^{s - 1} ||r^1  - r^0 ||_\infty   + \lambda ^{s - 2} ||r^1  - r^0 ||_\infty   +  \cdots
  +\\
  \quad \quad \quad \quad \quad \lambda ^t ||r^1  - r^0 ||_\infty   \\
  \quad \quad \quad \; \; \; \le ||r^1  - r^0 ||_\infty  \lambda ^t \sum\limits_{k = 0}^{s - t - 1} {\lambda ^k }   < ||r^1  - r^0 ||_\infty  \lambda ^t \sum\limits_{k = 0}^\infty  {\lambda ^k }  \\
  \quad \quad \quad \; \; \; = ||r^1  - r^0 ||_\infty  \lambda ^t \frac{1}{{1 - \lambda
  }} \\
  \quad \quad \quad \; \; \; \le ||r^1  - r^0 ||_\infty  \lambda ^N \frac{1}{{1 - \lambda }} \\ \quad \quad \quad \; \; \; \le \varepsilon,  \\
 \end{array}
 \]

\noindent where the first inequality holds by the triangle inequality,
and the second inequality is due to Lemma \ref{lem:converge}. Then, by
Cauchy convergence theorem \cite{74analysis}, we conclude that the
sequence $r _\alpha ^ k$ converges to a fixed point. For the
uniqueness, we prove it by contradiction. Suppose
Eq.~(\ref{eq:rankiter}) has at least two fixed points. Let $r ^{(1)}$
and $r ^{(2)}$ be two fixed points, and $M=|r_i ^ {(1)} - r_i ^
{(2)}|=||r ^{(1)} - r ^{(2)}||_\infty$. Then, we have

\[
\begin{array}{l}
 M = |\frac{1}{{|I_i| }}\sum\limits_{j \in I_i } {W_{ji} ((f(r^{(1)}))_j - (f(r^{(2)}))_j)} | \\
  \quad \;  \le \frac{1}{{|I_i| }}\sum\limits_{j \in I_i } {W_{ji} |((f(r^{(1)}))_j - (f(r^{(2)}))_j)|}  \\
  \quad \;  \le \frac{\lambda }{{|I_i| }}\sum\limits_{j \in I_i } {W_{ji} ||r^{(1)}  - r^{(2)} ||_\infty  }  \\
  \quad \;  \le \lambda ||r^{(1)}  - r^{(2)} ||_\infty = \lambda M. \\
 \end{array}
\]

\noindent Since $\lambda \in [0, 1)$, thus $M < M$, which is a
contradiction. This completes the proof. \eop
\end{myproof}

\stitle{The rate of the convergence: }We show that our algorithms will
converge in exponential rate by the following lemmas.

\begin{lemma}
\label{lem:convergerate} $|| r ^ \infty - r ^ k ||_\infty \le
\lambda ^k ||r ^ \infty - r ^0|| _\infty$.
\end{lemma}

\begin{myproof}
We prove the lemma by induction. For $k=1$, let $|r_i ^ \infty - r_i
^1| = || r ^ \infty - r ^ 1 ||_\infty$, then we have

\[
\begin{array}{l}
 |r_i^\infty   - r_i^1 | = |\frac{1}{{|I_i| }}\sum\limits_{j \in I_i } {W_{ji} ((f(r^0 ))_j - (f(r^\infty ))_j)} | \\
  \quad \quad \quad \quad \; \le \frac{1}{{|I_i| }}\sum\limits_{j \in I_i } {W_{ji} |(f(r^0))_j - (f(r^\infty ))_j)|}  \\
  \quad \quad \quad \quad \; \le \frac{\lambda }{{|I_i|
  }}\sum\limits_{j \in I_i } {W_{ji} ||r^\infty   - r^0 ||_\infty  }
\le \lambda ||r^\infty   - r^0 ||_\infty \\
 \end{array}
\]

\noindent The last inequality holds by the definition of
vector-valued contractive function. Suppose $k=t$, we have $ || r ^
\infty - r ^ t ||_\infty \le \lambda ^t ||r ^ \infty - r ^0||
_\infty$. Then, when $k=t+1$, for any node $u$ of the graph, we have

\[
\begin{array}{l}
 |r_u^\infty   - r_u^{t+1} | = |\frac{1}{{|I_u| }}\sum\limits_{j \in I_u } {W_{ju} ((f(r^t ))_j - (f(r^\infty))_j)} | \\
  \quad \quad \quad \quad \quad \; \le \frac{1}{{|I_u| }}\sum\limits_{j \in I_u } {W_{ju} |(f(r^t ))_j - (f(r^\infty))_j)|}  \\
  \quad \quad \quad \quad \quad \; \le \frac{\lambda }{{|I_u| }}\sum\limits_{j \in I_u } {W_{ju} ||r^\infty   - r^t ||_\infty  }  \\
  \quad \quad \quad \quad \quad \; \le \lambda ||r^\infty   - r^t
  ||_\infty
\le \lambda ^ {t+1} ||r ^ \infty - r ^0|| _\infty. \\
 \end{array}
\]

\noindent Thus, we have $||r^\infty   - r^t ||_\infty \le \lambda ^
{t+1} ||r ^ \infty - r ^0|| _\infty$. This completes the proof. \eop
\end{myproof}

\begin{lemma}
\label{lem:normbound} $|| r ^ a - r ^ b ||_\infty \le 1$.
\end{lemma}

\begin{myproof}
By definition, for any $t$, $f(r ^ t) \le e$ holds. Thus, we
conclude $|| r ^ a - r ^ b ||_\infty \le 1$. \eop
\end{myproof}

With the above lemma, we readily have the following corollary.

\begin{corollary}
  \label{cor:convergerate}
  $|| r ^ \infty - r ^ k ||_\infty \le \lambda ^k$.
\end{corollary}

By Corollary \ref{cor:convergerate}, our algorithms converge in
exponential rate. We can determine the maximal steps that are needed
for convergence. Assume $r_i$ is the true prestige score of node
$i$. Our goal is to show that after a particular number of iterations
$k$, the prestige score given by our algorithm converges to $r_i$ as
desired. Formally, for $\varepsilon \to 0$, let $|r_i - r_i ^ k| \le
\varepsilon $. By Corollary \ref{cor:convergerate}, we can set
\begin{equation}
k = \log _\lambda \varepsilon.
\end{equation}
This implies that the number of iterations $k$ is a very small
constant to guarantee convergence of our proposed algorithms. This
also leads to the linear time complexity of our algorithms as we will
discuss in Section \ref{subsec:complexity}.

\vspace*{-0.2cm}
\subsection{Instances of {\large $f(r)$}}
\label{subsec:specialinstance}

In this section, we first show that MB is a special instance of our
framework on unsigned trust networks. Then, based on our framework, we
present four new algorithms that can circumvent the existing problems
of MB. The proofs in this section are given in Appendix.

To show that MB on the unsigned trust network is a special instance of
our algorithm, we show that $f_{mb}(r)$ is a vector-valued contractive
function. The $f_{mb}(r)$ is defined by
\[(f_{mb}(r))_j
= \max \{ 0,\frac{1}{{2|O_j |}}\sum\limits_{i \in O_j } {(W_{ji}  -
r_i )\} }, \]
for $j=1,2, \cdots, n$. In particular, we have the following theorem.

\begin{theorem}
  \label{thm:mbinstance} For any $r \in \mathbb{R} ^n$, and $r \le e$,
  $f_{mb}$ is a vector-valued contractive function with the decay
  constant $\lambda = 1/2$ and $0 \le f_{mb} \le e$.
\end{theorem}

As analysis in the previous section, MB yields unreasonable bias
measurement and it is easy to be attacked by the adversarial nodes. In
the following, we propose four new algorithms that can tackle the
existing problems in MB.  Specifically, we give two classes of
vector-valued contractive functions: the $L_1$ distance based
vector-valued contractive functions and the $L_2$ distance based
vector-valued contractive functions. All functions can be served as
$(f(r^{k+1}))_j$ in Eq.~(\ref{eq:rankiter}). That is to say, all of
these functions can be used to measure the bias of the nodes.

\stitle{$L_1$ distance based contractive functions:} We present two
vector-valued contractive functions based on the $L_1$ distance
measure: $f_1(r)$ and $f_2(r)$. Specifically,

\begin{equation}
\label{eq:l1normiter} (f_1 (r))_j  = \frac{\lambda }{{|O_j
|}}\sum\limits_{i \in O_j } {|W_{ji}  - r_i |},
\end{equation}

\noindent
for all $j=1,2,\cdots, n$. In the following theorem, we show that
$f_1$ is a vector-valued contractive function.

\begin{theorem}
\label{thm:f1} For any $r \in \mathbb{R} ^n$, and $r \le e$, $f_1$
is a vector-valued contractive function with $0 \le f_1 \le e$.
\end{theorem}

Based on $f_1$, the bias of node $j$ is determined by the arithmetic
average of the differences between the trust scores given by node $j$
and the corresponding prestige scores of the outgoing neighbors of
node $j$. The rationale is that the nodes whose trust scores often
differ from those of other nodes will be assigned to high bias
scores. In $f_1$, the difference is measured by the $L_1$ distance,
thus we refer to this algorithm as the $L_1$ average \tw algorithm
($L_1$-AVG). The corresponding iterative system is given by

\begin{equation}
\label{eq:iterf1} \left\{
\begin{array}{l}
 r_i^{k + 1}  = \frac{1}{{|I_i| }}\sum\limits_{j \in I_i } {W_{ji} (1 - (f_1(r^k ))_j )}  \\
 (f_1 (r^{k + 1} ))_j  = \frac{\lambda }{{|O_j |}}\sum\limits_{i \in O_j } {|W_{ji}  - r_i^{k + 1} |}.  \\
 \end{array} \right.
\end{equation}

\noindent
It is important to note that, unlike MB, $L_1$-AVG uses the $L_1$
distance to measure the differences, thus the differences between the
trust score and the corresponding prestige score cannot be canceled
out. It therefore can readily prevent attacks from the adversarial
nodes that give the nodes with high prestige low trust scores and give
the nodes with low prestige high trust scores.
%
%
Table~\ref{tbl:l1avgbiasexample} shows the bias scores of the nodes
for the example in Fig.~\ref{fig:example} by $L_1$-AVG.  For fair
comparison with MB, we set $\lambda = 0.5$ in all of our algorithms in
this experiment. We can clearly see that node 5 achieves the highest
bias score, which conforms with our intuition. Also, we can observe
that $L_1$-AVG converges in 5 iterations, because the rate of
convergence of our framework is exponential.


\begin{table}[t]
\begin{center}
\caption{Bias scores by the $L_1$-AVG algorithm.} {\small
\begin{tabular}{|l|c|c|c|c|c|}
\hline
Iteration & node 1 & node 2 & node 3 & node 4 & node 5 \\
\hline
1 & 0.115 & 0.200 & 0.292 & 0.111 & 0.207 \\
2 & 0.005 & 0.130 & 0.137 & 0.060 & 0.220 \\
3 & 0.019 & 0.117 & 0.098 & 0.054 & 0.233 \\
4 & 0.018 & 0.113 & 0.089 & 0.054 & 0.237 \\
5 & 0.018 & 0.113 & 0.089 & 0.054 & 0.237 \\
\hline
\end{tabular}
} \label{tbl:l1avgbiasexample}
\end{center}\vspace*{-0.5cm}
\end{table}

The second $L_1$-distance based vector-valued contractive function
is defined by

\begin{equation}
\label{eq:maxl1iter} (f_2 (r))_j  = \lambda \mathop {\max}\limits_{i
\in O_j } |W_{ji}  - r_i |,
\end{equation}

\noindent
for all $j=1,2,\cdots, n$. Below, we show that $f_2$ is a
vector-valued contractive function.

\begin{theorem}
\label{thm:f2} For any $r \in \mathbb{R} ^n$, and $r \le e$, $f_2$
is a vector-valued contractive function with $0 \le f_2 \le e$.
\end{theorem}

In $f_2$, since the bias of node $j$ is determined by the maximal
difference between the trust scores given by node $j$ and the
corresponding prestige score of the outgoing neighbors of node $j$,
we refer to this algorithm as the $L_1$ maximal \tw algorithm
($L_1$-MAX). The corresponding iterative system is as follows.

\begin{equation}\label{eq:iterf2}\left\{
\begin{array}{l}
 r_i^{k + 1}  = \frac{1}{{|I_i|}}\sum\limits_{j \in I_i } {W_{ji} (1 - (f_2(r^k ))_j )}  \\
 (f_2 (r^{k + 1} ))_j  = \lambda \mathop {\max }\limits_{i \in O_j } |W_{ji}  - r_i^{k + 1} |. \\
 \end{array} \right.
 \end{equation}

\noindent
With Eq.~(\ref{eq:maxl1iter}), we can see that $L_1$-MAX punishes the
biased nodes more heavily than $L_1$-AVG, as it takes the maximal
difference to measure the bias.  In other words, in $L_1$-MAX, the
node that only gives one unreasonable trust score will get high bias
score. Like $L_1$-AVG, $L_1$-MAX can also prevent attacks from the
adversarial nodes that give the nodes with high prestige low trust
scores, and give the nodes with low prestige high scores.
%
%
Table \ref{tbl:l1maxbiasexample} shows the bias scores of the nodes
for the example in Fig.~\ref{fig:example} by $L_1$-MAX. We can see
that node 5 gets the highest bias score as desired. $L_1$-MAX
converges in 5 iterations, because the rate of convergence of our
framework is exponential.

\begin{table}[t]
\begin{center}
\caption{Bias scores by the $L_1$-MAX algorithm.} {\small
\begin{tabular}{|l|c|c|c|c|c|}
\hline
Iteration & node 1 & node 2 & node 3 & node 4 & node 5 \\
\hline
1 & 0.115 & 0.343 & 0.343 & 0.165 & 0.407 \\
2 & 0.000 & 0.215 & 0.215 & 0.050 & 0.311 \\
3 & 0.020 & 0.179 & 0.179 & 0.061 & 0.289 \\
4 & 0.017 & 0.169 & 0.169 & 0.065 & 0.285 \\
5 & 0.017 & 0.169 & 0.169 & 0.065 & 0.285 \\
\hline
\end{tabular}
} \label{tbl:l1maxbiasexample}
\end{center}\vspace*{-0.5cm}
\end{table}

\stitle{$L_2$ distance based contractive functions:} We propose two
contractive functions based on the square of $L_2$ distance measure.
For convenience, we refer to these functions as $L_2$ distance based
contractive functions. Since the $L_2$ distance based algorithms are
defined in a similar fashion as the $L_1$ distance based algorithms,
we omit explanation unless necessary. The first $L_2$ distance based
contractive function is given by the following equation.

\begin{equation}
\label{eq:qudraiter} (f_3 (r))_j  = \frac{\lambda }{{2|O_j
|}}\sum\limits_{i \in O_j } {(W_{ji}  - r_i )^2 },
\end{equation}

\noindent
for all $j=1,2,\cdots, n$. We can also prove the $f_3$ is a
vector-valued contractive function.

\begin{theorem}
\label{thm:f3} For any $r \in \mathbb{R} ^n$, and $r \le e$,
$f_3(r)$ is a vector-valued contractive function with $0 \le f_3(r)
\le e$.
\end{theorem}

Similarly, in $f_3$, the bias of node $j$ is determined by the
arithmetic average of the difference between the trust scores given
by node $j$ and the corresponding prestige score of the outgoing
neighbors of node $j$. However, unlike $f_1$ and $f_2$, in $f_3$,
the difference is measured by the square of $L_2$ distance. Thus, we
refer to this algorithm as the $L_2$ average \tw algorithm
($L_2$-AVG). The corresponding iterative system is

\begin{equation}
\left\{ \begin{array}{l}
 r_i^{k + 1}  = \frac{1}{{|I_i|}}\sum\limits_{j \in I_i } {W_{ji} (1 - (f_3(r^k ))_j )}  \\
 (f_3 (r^{k + 1} ))_j  = \frac{\lambda }{{2|O_j |}}\sum\limits_{i \in O_j } {(W_{ji}  - r_i^{k + 1} )^2 }.  \\
 \end{array} \right.
\end{equation}

The second $L_2$ distance based vector-valued contractive function
is defined by

\begin{equation}
\label{eq:maxl2iter}  (f_4 (r))_j  = \frac{\lambda }{2}\mathop {\max
}\limits_{i \in O_j } (W_{ji}  - r_i )^2,
\end{equation}

\noindent
for all $j=1,2,\cdots, n$. Likewise, we have the following theorem.

\begin{theorem}
\label{thm:f4} For any $r \in \mathbb{R} ^n$, and $r \le e$,
$f_4(r)$ is a vector-valued contractive function with $0 \le f_4(r)
\le e$.
\end{theorem}

The corresponding iterative system is

\begin{equation}
\left\{ \begin{array}{l}
 r_i^{k + 1}  = \frac{1}{{|I_i| }}\sum\limits_{j \in I_i } {W_{ji} (1 - (f_4(r^k ))_j )}  \\
 (f_4 (r^{k + 1} ))_j  = \frac{\lambda }{2}\mathop {\max }\limits_{i \in O_j } (W_{ji}  - r_i^{k+1} )^2 . \\
 \end{array} \right.
\end{equation}

\noindent
Similar to $L_1$-MAX, we refer to this algorithm as the $L_2$ maximal
\tw algorithm ($L_2$-MAX).

We depict the prestige scores by different algorithms in Table
\ref{tbl:prestigeexample}. We can observe that the rank of the
prestige scores by our algorithms is the same as the rank by AA
(Arithmetic average) algorithm in Fig.~\ref{fig:example}, and also it
is strongly correlated to MB. Note that all of our algorithms give
zero prestige score to node 2, as node 2 obtains zero trust score from
his/her incoming neighbors.

\begin{table}[t]
\begin{center}
\caption{Prestige scores by different algorithms.} {\small
\begin{tabular}{|l|c|c|c|c|c|}
\hline
Algorithm & node 1 & node 2 & node 3 & node 4 & node 5 \\
\hline
AA & 0.567 & 0.000 & 0.433 & 0.600 & 0.350 \\
HITS & 1.000 & 0.000 & 0.401 & 0.391 & 0.027 \\
PageRank & 0.224 & 0.030 & 0.305 & 0.141 & 0.300 \\
MB & 0.532 & 0.000 & 0.433 & 0.523 & 0.350 \\
$L_1$-AVG & 0.502 & 0.000 & 0.352 & 0.541 & 0.336  \\
$L_1$-MAX & 0.461 & 0.000 & 0.331 & 0.492 & 0.335  \\
$L_2$-AVG & 0.558 & 0.000 & 0.416 & 0.594 & 0.349  \\
$L_2$-MAX & 0.556 & 0.000 & 0.414 & 0.591 & 0.348  \\
\hline
\end{tabular}
} \label{tbl:prestigeexample}
\end{center}\vspace*{-0.5cm}
\end{table}

\subsection{Complexity of the proposed algorithms}
\label{subsec:complexity}

In this section, we analyze the time and space complexities of
$L_1$-AVG. For the other algorithms, it is not hard to show that the
time and space complexities are the same as $L_1$-AVG.
First, the time complexity for computing the prestige score of node
$i$ in one iteration is $O(|\bar I||\bar O|)$, where $|\bar I|$ and
$|\bar O|$ denote the average in-degree and out-degree of all nodes
respectively. The amortized time complexity in one iteration is
$O(m)$, where $m$ denotes the number of edges in the graph.
Therefore, the total time complexity of $L_1$-AVG is $O(km)$, where
$k$ denotes the number of iterations that are needed to guarantee
convergence. As analyzed in Section \ref{subsec:analysis}, $k$ is a
very small constant. And $k=15$ can guarantee the algorithms
converge as shown in our experiments. The analysis implies that the
time complexity of our algorithms is linear w.r.t.\ the size of the
graph. Second, we only need to store the graph, the prestige vector
($r$), and the contractive function $f(r)$, thus the space
complexity is $O(m+n)$.  In summary, our algorithms have linear time
and space complexities, thereby they can be scalable to large
graphs.

\subsection{Generalizing to signed trust networks}
\label{subsec:genesignednet}

Our algorithms can be generalized to signed trust networks. In signed
trust networks, there exist two types of edges: the positive edge and
the negative edge. In other words, the weights of positive (negative)
edges are positive (negative). In practice, many trust networks, such
as Slashdot and Epinions, are signed trust networks, where the
negative edges signify distrust. Without loss of generality, we assume
that the weights of the edges have been scaled into [-1, 1].  It is
easy to verify that Lemma \ref{lem:converge}, Theorem
\ref{thm:converge}, and Lemma \ref{lem:convergerate} still hold in
signed trust networks. Therefore, our iterative algorithms converge
into a unique fixed point in the context of signed trust networks.
Moreover, the rate of convergence is exponential. Notice that this
result holds if the function $f$ is a vector-valued contractive
function. In signed trust networks, it is easy to check that the
functions $f_1$ and $f_2$ are still the vector-valued contractive
functions, but the $f_3$ and $f_4$ are not. However, we can readily
modify them to the vector-valued contractive functions, which are
denoted by $f_3 ^*$ and $f_4 ^*$ respectively, by adjusting the decay
constant. Specifically, we have

\[
\begin{array}{l}
 (f_3^* (r))_j  = \frac{\lambda }{{4|O_j |}}\sum\limits_{i \in O_j } {(W_{ji}  - r_i )^2 }  \\
 (f_4^* (r))_j  = \frac{\lambda }{4}\mathop {\max }\limits_{i \in O_j } (W_{ji}  - r_i )^2.  \\
 \end{array}
\]

\noindent
It is easy to verify that $f_3^* (r)$ and $f_4^* (r)$ are
vector-valued contractive functions in signed trust networks.

\section{Experiments}
\label{sec:experiments}

In this section, we evaluate the effectiveness, robustness and
scalability of our algorithms.

\subsection{Experimental setup}

\stitle{Datasets}: We conduct our experiments on five real datasets.
(1) Kaitiaki dataset: We collect the Kaitiaki dataset from Trustlet
(\url{www.trustlet.org}). This dataset is a trust network dataset,
where the trust statements are weighted at four different levels (0.4,
0.6, 0.8, and 1.0). (2) Epinions dataset: We download it from Stanford
network analysis data collections (\url{http://snap.stanford.edu}). It
is a signed trust network dataset, where the users can trust or
distrust the other users. (3) Slashdot datasets: we collect three
different datasets from Stanford network analysis data
collections. All of these three datasets are signed trust networks,
where the users can give trust or distrust scores to the others. Table
\ref{tbl:data} summarizes the detailed statistical information of the
datasets.

\begin{table}[t]
\begin{center}
\caption[]{Summary of the datasets} \label{tbl:data} {\small
\begin{tabular}{|l|c|c|c|}
\hline
Name & Nodes & Edges & Ref.\\
\hline
Kaitiaki & 64 & 178 & - \\
Epinions & 131,828 & 841,372 & \cite{10chisignedgraph} \\
Slashdot1 & 77,350 & 516,575 & \cite{10chisignedgraph}\\
Slashdot2 & 81,867 & 545, 671 & \cite{10chisignedgraph}\\
Slashdot3 & 82,140 & 549,202 & \cite{10chisignedgraph}\\
\hline
\end{tabular}}
\end{center}\vspace*{-0.8cm}
\end{table}

\stitle{Parameter settings and experimental environment:} We set the
decay constant $\lambda=0.5$ for a fair comparison with MB. For the
decay constant of the PageRank algorithm, we set it to 0.85, as it is
widely used in web search. All the experiments are conducted on a
Windows Server 2008 with 4x6-core Intel Xeon 2.66 Ghz CPU, and 128
memory. All algorithms are implemented by MATLAB 2010a and Visual C++
6.0.

\subsection{Experimental results}

\stitle{Comparison of bias score}: Here we compare the bias scores
by our algorithms with the bias scores by MB. First, we use the
variance of the trust scores given by node $i$ to measure the bias
of the node $i$, as used in \cite{11wwwbiasrank}. Specifically, we
define the variance as follows:
%
\begin{equation}\label{eq:variance} {\mathop{\rm var}} (i)
= \frac{1}{{|O_i |}}\sum\limits_{j \in O_i } {(W_{ij}  - \bar r_j
)^2 },
\end{equation}
%
where $\bar r_j = \frac{1}{{|I_j |}}\sum\limits_{i \in I_j }
{W_{ij}}$. Second, we rank the nodes by their variance and use this
rank as the ``ground truth''. Note that there is no ground truth for
the bias score of the nodes in any datasets. We use the variance as
the ground truth. The reason is twofold. On one hand, the variance is
an intuitive metric for measuring the bias of the node, and the node
having a larger variance implies that the node has a larger bias
score. On the other hand, the variance has been successfully used for
analyzing the bias of the node in trust networks
\cite{11wwwbiasrank}.
Third, we rank the nodes by their bias scores obtained by our
algorithms and obtained by MB, respectively. Specifically, for MB, we
rank the nodes by the absolute value of the bias scores ($|b_i|$ in
Eq.~(\ref{def:mbbias})). Finally, we compare our algorithms with MB in
terms of AUC (the area under the ROC curve) \cite{83auc} and Kendall
Tau \cite{38kendalltau} metric, where the AUC metric is used to
evaluate the top-K rank (in our experiments, we consider the top-5\%
nodes) and the Kendall Tau metric is employed to evaluate the rank
correlation between the rank by the proposed algorithms and the ground
truth.

Table \ref{tbl:biasauc} and Table \ref{tbl:biaskendalltau} show the
comparison of bias by our algorithms and MB under AUC and Kendall Tau
metric, respectively. From Table \ref{tbl:biasauc}, we can see that
$L_1$-AVG and $L_2$-AVG achieve the best performance. In signed trust
networks, the performance of our algorithms are significantly better
than MB. For example, $L_2$-AVG boosts AUC over MB by 4.7\%, 11\%,
9.9\%, and 9.7\% in Epinions, Slashdot1, Slashdot2 and Slashdot3,
respectively. The results indicate that our algorithms are more
effective than MB for computing the bias of the nodes. This is because
the bias measurements of our algorithms are more reasonable than the
bias measurement of MB. Interestingly, $L_1$-AVG and $L_2$-AVG achieve
the same performance under the AUC metric. In general, $L_1$-AVG and
$L_2$-AVG outperform $L_1$-MAX and $L_2$-MAX in our datasets.  From
Table \ref{tbl:biaskendalltau}, we can observe that all the algorithms
exhibit positive correlation to the ground truth. $L_2$-AVG achieves
the best performance in Kaitiaki, Epinions, Slashdot1, and Slashdot3
datasets, while in Slashdot2 dataset $L_1$-AVG achieves the best
performance. It is important to note that all of our algorithms
significantly outperform MB in signed networks. For instance,
$L_2$-AVG improves Kendall Tau over MB by 11.9\%, 6.8\%, 10.1\%,
12.3\%, and 13.9\% in Kaitiaki, Epinions, Slashdot1, Slashdot2 and
Slashdot3, respectively. The results further confirm that our
algorithms are more effective than MB for computing the bias of the
node in trust networks.

\begin{table}[t]
\begin{center}
\caption[]{
 Comparison of bias
 by our algorithms and MB algorithm
under AUC metric (top 5\% nodes of the dataset).}
\label{tbl:biasauc} {\small
\begin{tabular}{|l|c|c|c|c|c|}
\hline
Datasets & $L_1$-AVG & $L_1$-MAX & $L_2$-AVG & $L_2$-MAX & MB \\
\hline
Kaitiaki & \textbf{1.000} & 0.937 & \textbf{1.000} & 0.925 & \textbf{1.000} \\
Epinions & \textbf{0.994} & 0.982 & \textbf{0.994} & 0.982 & 0.949 \\
Slashdot1 & \textbf{0.993} & 0.970 & \textbf{0.993} & 0.970 & 0.895 \\
Slashdot2 & \textbf{0.992} & 0.975 & \textbf{0.992} & 0.975 & 0.903 \\
Slashdot3 & \textbf{0.992} & 0.975 & \textbf{0.992} & 0.975 & 0.903 \\
\hline
\end{tabular}}
\end{center}
\vspace*{-0.8cm}
\end{table}

\begin{table}[t]
\begin{center}
\caption[]{
 Comparison of bias
 by our algorithms and MB algorithm
under Kendall Tau metric.} \label{tbl:biaskendalltau} {\small
\begin{tabular}{|l|c|c|c|c|c|}
\hline
Datasets & $L_1$-AVG & $L_1$-MAX & $L_2$-AVG & $L_2$-MAX & MB \\
\hline
Kaitiaki & 0.728 & 0.713 & \textbf{0.812} & 0.709 & 0.726 \\
Epinions & 0.781 & 0.754 & \textbf{0.783} & 0.754 & 0.733 \\
Slashdot1 & 0.811 & 0.776 & \textbf{0.812} & 0.776 & 0.734 \\
Slashdot2 & \textbf{0.722} & 0.688 & 0.721 & 0.688 & 0.642 \\
Slashdot3 & 0.820 & 0.787 & \textbf{0.821} & 0.787 & 0.721 \\
\hline
\end{tabular}}
\end{center}
\vspace*{-0.8cm}
\end{table}

\begin{figure}[t]
\begin{center}
\includegraphics[width=0.9\hsize]{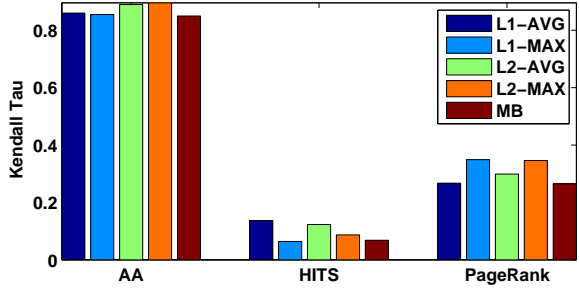}
 \vspace*{-1em}
\caption{
 Comparison of prestige
 by our algorithms and MB algorithm
on Kaitiaki dataset.} \label{fig:kprestige}
\end{center}
\vspace*{-0.5cm}
\end{figure}

\begin{figure}[t] 
\begin{center}
\includegraphics[width=\hsize]{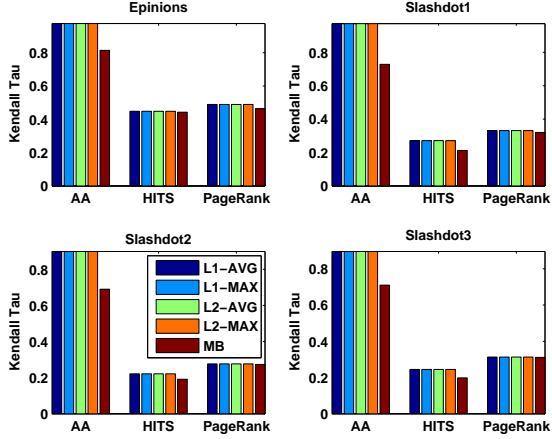}
\vspace*{-3em} \caption{Comparison of prestige
 by our algorithms and MB algorithm
on signed trust networks.} \label{fig:snprestige}
\end{center}
\vspace*{-0.8cm}
\end{figure}

\stitle{Comparison of prestige score:} We compare the prestige scores
by our algorithms with those by MB. Specifically, we use the
arithmetic average (AA), HITS \cite{99jacmhits}, and PageRank
\cite{pagerank} as the baselines, and then compare the rank
correlation between the rank by our algorithms (here we rank the nodes
according to their prestige scores) and the rank by the baselines
using Kendall Tau metric.  Here, AA ranks the nodes by the average
trust scores obtained from the incoming neighbors, and HITS ranks the
nodes by their authority scores. In signed trust networks, we remove
the signed edges for HITS and PageRank, as these algorithms cannot
work on signed trust networks directly.  Similar processing has been
used in \cite{11wwwbiasrank}.  Fig.~\ref{fig:kprestige} and
Fig.~\ref{fig:snprestige} depict the comparison of prestige score by
our algorithms and MB on Kaitiaki and signed trust networks,
respectively.

From Fig.~\ref{fig:kprestige}, we can clearly see that our algorithms
achieve the best rank correlation to AA. By comparing the Kendall Tau
between different algorithms (our algorithms and MB) and HITS, we find
that $L_1$-AVG achieves the best rank correlation.  However, by
comparing the Kendall Tau between different algorithms and PageRank,
we clearly find that $L_1$-MAX achieves the best rank correlation.
From Fig.~\ref{fig:snprestige}, we can also observe that our
algorithms achieve the best rank correlation to AA. By comparing the
rank correlation between different algorithms and HITS/PageRank, we
find that our algorithms are slightly better than MB on the signed
trust network datasets.  These results suggest that our algorithms are
more effective to measure the prestige of the nodes than MB.
Interestingly, all of our algorithms achieve the same performance in
signed trust networks.

\stitle{Robustness testing:} To evaluate the robustness of our
algorithms, we first add some noisy data into the original datasets.
Specifically, we randomly select some nodes as the spamming nodes, and
then modify the trust scores given by the spamming nodes. For the
spamming nodes, we randomly give high trust score to his/her
out-neighbors whose average trust score is low, and randomly give low
trust score to his/her outgoing neighbors whose average trust score is
high. Second, we perform our algorithms and MB on both original and
noisy datasets, and then calculate the Kendall Tau for each
algorithm. Here the Kendall Tau is computed on two ranks that are
yielded by an algorithm on the original datasets and the noisy
datasets, respectively.  Finally, we compare the Kendall Tau among all
algorithms.  Intuitively, the larger Kendall Tau the algorithm
achieves, the more robust the algorithm is.

\begin{figure}[t] 
\begin{center}
\includegraphics[width=0.9\hsize]{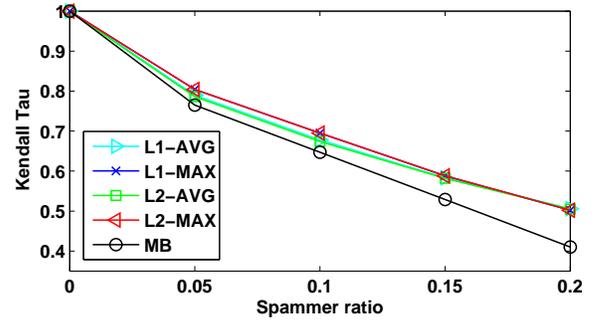}
\vspace*{-1em}
\caption{Robustness of bias
 by our algorithms and MB
 algorithm
on Epinions dataset.} \label{fig:biasrobust}
\end{center}
\vspace*{-0.5cm}
\end{figure}

\begin{figure}[t] 
\begin{center}
\includegraphics[width=0.9\hsize]{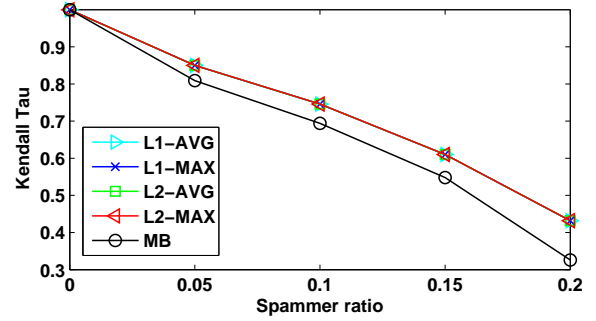}
\vspace*{-1em} \caption{Robustness of prestige
 by our algorithms and
 MB algorithm
on Epinions dataset.} \label{fig:prestigerobust}
\end{center}
\vspace*{-0.8cm}
\end{figure}

We test our algorithms and MB on both original and noisy datasets with
5\% to 20\% spamming ratio.  Fig.~\ref{fig:biasrobust} and
Fig.~\ref{fig:prestigerobust} describe the robustness of the bias and
the prestige of the algorithms by Kendall Tau vs. spamming ratio on
Epinions dataset, respectively.  Similar results can be obtained from
other datasets. From Fig.~\ref{fig:biasrobust} and
Fig.~\ref{fig:prestigerobust}, we can clearly see that all of our
algorithms are significantly more robust than MB. For the bias,
$L_2$-MAX achieves the best robustness, followed by the $L_1$-MAX,
$L_2$-AVG, $L_1$-AVG, and then MB. For the prestige, all of our
algorithms achieve the same robustness, and are significantly more
robust than MB. These results confirm our analysis in Section
\ref{sec:algorithms}. Moreover, the gap of robustness between our
algorithms and MB increases as the spamming ratio increases, which
suggests that our algorithms are more effective than MB
on the datasets with high spamming ratio. In general, the robustness
of the algorithms decrease as the spamming ratio increases.

\begin{figure}[t]
\begin{center}
\includegraphics[width=0.9\hsize]{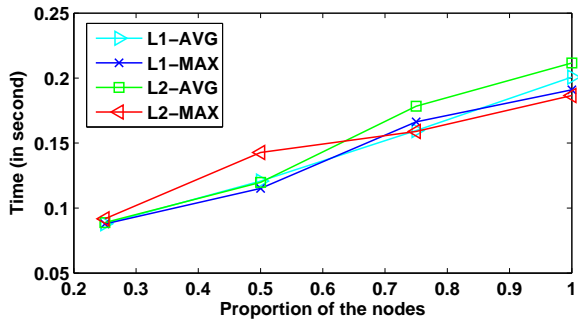}
\vspace*{-1em}
\caption{Scalability of the proposed algorithms.}
\label{fig:scale}
\end{center}
\vspace*{-0.5cm}
\end{figure}

\stitle{Scalability:} We evaluate the scalability of our algorithms
on the Epinions dataset. Similar results can be obtained from other
datasets. For evaluating the scalability, we first generate three
subgraphs in terms of the following rule. First, we randomly select
25\% nodes and the corresponding edges of the original graph as the
first dataset, and then add another 25\% nodes to generate the
second dataset, and then based on the second dataset, we add another
25\% nodes to generate the third dataset. Then, we perform our
algorithms on this three datasets and the original dataset.
Fig.~\ref{fig:scale} shows our results. From Fig.~\ref{fig:scale},
we can clearly see that our algorithms scales linearly w.r.t.\ the
size of the graph. This result conforms with our complexity analysis
in Section \ref{subsec:complexity}.

\begin{figure}[t] 
\begin{center}
\includegraphics[width=\hsize]{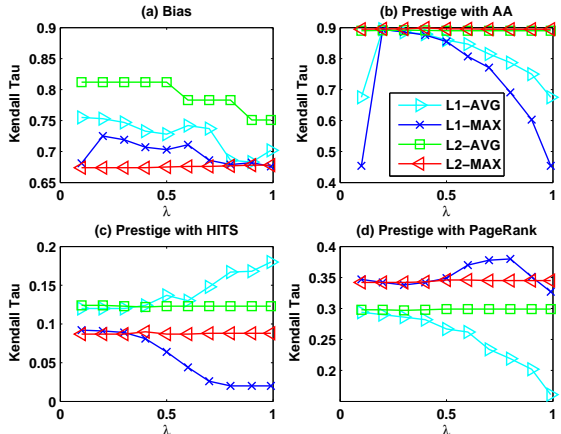}
\vspace*{-3em}
\caption{Effect of $\lambda$.}
\label{fig:labmdaeffect}
\end{center}
\vspace*{-0.8cm}
\end{figure}

\stitle{Effect of parameter $\lambda$}: We discuss the effectiveness
of parameter $\lambda$ in our algorithms on Kaitiaki dataset.
Similar results can be observed from other datasets.
Fig.~\ref{fig:labmdaeffect} shows the effectiveness of our
algorithms w.r.t.\ $\lambda$, where the effectiveness is measured by
the rank correlation between our algorithms and the baselines using
the Kendall Tau metric. Specifically, Fig.~\ref{fig:labmdaeffect}(a)
depicts the bias correlation between our algorithms and the
\emph{variance} based algorithm (Eq.~(\ref{eq:variance})) under
various $\lambda$, while Figs.~\ref{fig:labmdaeffect}(b), (c), and
(d) show the prestige correlation between our algorithms and AA,
HITS, and PageRank under different $\lambda$, respectively. From
Fig.~\ref{fig:labmdaeffect}(a), we find that $L_2$-MAX is quite
robust w.r.t.\ $\lambda$, while the performance of $L_2$-AVG
decreases as $\lambda$ increases. In addition, we find that
$L_1$-AVG and $L_1$-MAX are slightly sensitive w.r.t.\ $\lambda$,
because the differences between the maximal and minimal bias
correlation of these two algorithms do not exceed 0.1. For the
prestige scores (Figs.~\ref{fig:labmdaeffect}(b), (c), and (d)), we
can clearly see that $L_2$-AVG and $L_2$-MAX are more robust w.r.t.\
$\lambda$, whereas $L_1$-AVG and $L_1$-MAX are sensitive w.r.t.\
$\lambda$. For instance, consider the prestige correlation with
PageRank (Fig.~\ref{fig:labmdaeffect}(d)), we can observe that the
performance of $L_1$-AVG decreases as $\lambda$ increases.  However,
the performance of $L_1$-MAX increases as $\lambda$ increases when
$\lambda \le 0.8$, and otherwise it decreases as $\lambda$
increases. To summarize, the $L_2$ distance based algorithms are
more robust w.r.t.\ the parameter $\lambda$ than the $L_1$ distance
based algorithms.

\vspace*{-0.2cm}
\section{Conclusion}
\label{sec:concl}

In this paper, we propose a framework of algorithms to calculate the
bias and prestige of the nodes in trust networks. Our framework is
general and includes a rich family of algorithms. We show that the
MB algorithm \cite{11wwwbiasrank} is a special case of our framework
on unsigned trust networks. Furthermore, we propose four new
algorithms based on our framework to circumvent the existing
problems of the MB algorithm. Our algorithms are more effective and
more robust than the MB algorithm. In addition, all of our
algorithms are scalable to handle large datasets, as they have
linear time and space complexities w.r.t.\ the size of the network.
Extensive experiments on five real datasets demonstrate the
effectiveness, robustness, and scalability of the proposed
algorithms. Future work includes developing new vector-valued
contractive functions and generalizing the proposed methods to
time-evolving trust networks.


\vspace*{-0.2cm}

\bibliographystyle{abbrv}
\bibliography{wwwrank}

\vspace*{-0.3cm}
\section*{Appendix} \small
\stitle{Proof of Theorem \ref{thm:mbinstance}:} For any $r, s \in
\mathbb{R} ^n$ and $j$, let \[
\begin{array}{l}
 \Delta _j  = |(f_{mb} (r))_j  - (f_{mb} (s))_j | \\
 \quad \; \; = |\max \{ 0,\frac{1}{{2|O_j |}}\sum\limits_{i \in O_j } {(W_{ji}  - r_i )} \}  - \\
 \quad \quad \; \; \max \{ 0,\frac{1}{{2|O_j |}}\sum\limits_{i \in O_j } {(W_{ji}  - s_i )} \} |. \\
 \end{array}
\] Consider the following four cases: \\
(A) $\frac{1}{{2|O_j |}}\sum\limits_{i \in O_j } {(W_{ji}  - r_i )}
\le 0$ and $ \frac{1}{{2|O_j |}}\sum\limits_{i \in O_j } {(W_{ji}  -
s_i )}  \le
0$. Obviously, $ \Delta _j  = 0 \le \frac{1}{2}||r - s||_\infty$. \\
(B) $ \frac{1}{{2|O_j |}}\sum\limits_{i \in O_j } {(W_{ji}  - r_i )}
\ge 0$ and $ \frac{1}{{2|O_j |}}\sum\limits_{i \in O_j } {(W_{ji} -
s_i )}  \ge 0$. We have
\[
\begin{array}{l}
 \Delta _j  = |\frac{1}{{2|O_j |}}\sum\limits_{i \in O_j } {(s_i  - r_i )} | \\
 \quad \; \; \le \frac{1}{{2|O_j |}}\sum\limits_{i \in O_j } {|s_i  - r_i |}  \\
 \quad \; \;  \le \frac{1}{{2|O_j |}}\sum\limits_{i \in O_j } {||r - s||_\infty} \\
 \quad \; \;  = \frac{1}{2}||r - s||_\infty.   \\
 \end{array}
\]
(C) $\frac{1}{{2|O_j |}}\sum\limits_{i \in O_j } {(W_{ji}  - r_i )}
\ge 0$ and $ \frac{1}{{2|O_j |}}\sum\limits_{i \in O_j } {(W_{ji} -
s_i )}  \le 0$. By $ \frac{1}{{2|O_j |}}\sum\limits_{i \in O_j }
{(W_{ji} - s_i )}  \le 0$, we have $ \sum\limits_{i \in O_j }
{W_{ji} }  \le \sum\limits_{i \in O_j } {s_i }$. Then, we have
\[
\begin{array}{l}
 \Delta _j  = \frac{1}{{2|O_j |}}\sum\limits_{i \in O_j } {(W_{ji}  - r_i )} \\
 \quad \; \; \le \frac{1}{{2|O_j |}}\sum\limits_{i \in O_j } {(s_i  - r_i )}  \\
  \quad \; \; \le \frac{1}{{2|O_j |}}\sum\limits_{i \in O_j } {|s_i  - r_i |}  \\
  \quad \; \; \le \frac{1}{2}||r - s||_\infty.   \\
 \end{array}
\]
(D) $\frac{1}{{2|O_j |}}\sum\limits_{i \in O_j } {(W_{ji}  - r_i )}
\le 0$ and $ \frac{1}{{2|O_j |}}\sum\limits_{i \in O_j } {(W_{ji} -
s_i )}  \ge 0$. Similar to the case (3), we have $ \Delta _j  \le
\frac{1}{2}||r - s||_\infty$. \\
To summarize, for any $j$, we have $ \Delta _j  \le \frac{1}{2}||r -
s||_\infty$. Hence, $f_{mb}$ is a vector-valued contractive function
with $\lambda = 1/2$. Since $0 \le W_{ji} \le 1$ and $r \le e$, thus
$0 \le f_{mb} \le e$. This completes the proof. \eop

\stitle{Proof of Theorem \ref{thm:f1}:} For any $r, s \in \mathbb{R}
^n$, we have
\[\begin{array}{l}
 |(f_1(r))_j - (f_1(s))_j| \\
 = |\frac{\lambda }{{|O_j |}}\sum\limits_{i \in O_j } {|W_{ji}  - r_i |}  - \frac{\lambda }{{|O_j |}}\sum\limits_{i \in O_j } {|W_{ji}  - s_i |} | \\
  = \frac{\lambda }{{|O_j |}}|\sum\limits_{i \in O_j } {(|W_{ji}  - r_i | - |W_{ji}  - s_i |} )| \\
  \le \frac{\lambda }{{|O_j |}}\sum\limits_{i \in O_j } {|r_i  - s_i |}  \\
  \le \lambda ||r - s||_\infty   \\
 \end{array}\] Since $0\le r \le e$, $0 \le W_{ji} \le 1$ and $0 \le \lambda < 1$,
thus $0 \le f_1 \le e$. \eop

\stitle{Proof of Theorem \ref{thm:f2}:} For any $r, s \in \mathbb{R}
^n$, let $|W_{ju}  - r_u |\mathop { = \max }\limits_{i \in O_j }
|W_{ji} - r_i |$, and $|W_{jv}  - s_v|\mathop { = \max }\limits_{i
\in O_j } |W_{ji}  - s_i |$, then we have
\[\begin{array}{l}
 |(f_2 (r ))_j - (f_2 (s ))_j| \\
 = |\lambda \mathop {\max }\limits_{i \in O_j } |W_{ji}  - r_i | - \lambda \mathop {\max }\limits_{i \in O_j } |W_{ji}  - s_i || \\
  \le \lambda \max \{ ||W_{ju}  - r_u | - |W_{ju}  - s_u ||,||W_{jv}  - r_v | - |W_{jv}  - s_v ||\}  \\
  \le \lambda \max \{ |r_u  - s_u |,|r_v  - s_v |\}  \\
  \le \lambda ||r - s||_\infty   \\
 \end{array}\] Since $0\le r \le e$, $0 \le W_{ji} \le 1$ and $0 \le \lambda < 1$,
thus $0 \le f_2 \le e$. \eop

\stitle{Proof of Theorem \ref{thm:f3}:} For any $r, s \in
\mathbb{R}$, and $r \le e, s\le e$, we have
\[
\begin{array}{l}
 |(f_3 (r))_j - (f_3 (s))_j| \\
 = |\frac{\lambda }{{2|O_j |}}\sum\limits_{i \in O_j } {(W_{ji}  - r_i )^2 }  - \frac{\lambda }{{2|O_j |}}\sum\limits_{i \in O_j } {(W_{ji}  - s_i )^2 } | \\
  \le \frac{\lambda }{{2|O_j |}}\sum\limits_{i \in O_j } {|(W_{ji}  - r_i )^2  - (W_{ji}  - s_i )^2 |}  \\
  = \frac{\lambda }{{2|O_j |}}\sum\limits_{i \in O_j } {|(s_i  - r_i )(2W_{ji}  - r_i  - s_i )|}  \\
  \le \frac{\lambda }{{|O_j |}}\sum\limits_{i \in O_j } {|s_i  - r_i |}  \\
  \le \lambda ||r - s||_\infty  \\
 \end{array}
\] Since $0\le r \le e$, $0 \le W_{ji} \le 1$ and $0 \le \lambda < 1$,
thus $0 \le f_3 \le e$. \eop

\stitle{Proof of Theorem \ref{thm:f4}:} For any $r, s \in
\mathbb{R}$, and $r \le e, s\le e$, let $ (W_{ju} - r_u )^2  =
\mathop {\max }\limits_{i \in O_j } (W_{ji}  - r_i )^2 $, and $
(W_{jv}  - s_v )^2  = \mathop {\max }\limits_{i \in O_j } (W_{ji}  -
s_i )^2$, then we have
\[
\begin{array}{l}
 |(f_4 (r))_j - (f_4 (s))_j| \\
 = |\frac{\lambda }{2}\mathop {\max }\limits_{i \in O_j } (W_{ji}  - r_i )^2  - \frac{\lambda }{2}\mathop {\max }\limits_{i \in O_j } (W_{ji}  - s_i )^2 | \\
  \le \frac{\lambda }{2}\max \{ |(W_{ju}  - r_u )^2  - (W_{ju}  - s_u )^2 |, \\
  \quad \quad \quad \quad|(W_{jv}  - s_v )^2  - (W_{jv}  - r_v )^2 |\}  \\
  = \frac{\lambda }{2}\max \{ |(s_u  - r_u )(2W_{ju}  - r_u  - s_u )|,\\
  \quad \quad \quad \quad|(s_v  - r_v )(2W_{jv}  - r_v  - s_v )|\}  \\
  \le \lambda \max \{ |s_u  - r_u |,|s_v  - r_v |\}  \\
  \le \lambda ||r - s||_\infty  \\
 \end{array}
\] Since $0\le r \le e$, $0 \le W_{ji} \le 1$ and $0 \le \lambda < 1$,
thus $0 \le f_4 \le e$. \eop

\end{document}